\documentclass[11pt]{article}
\usepackage{graphics,cite,amssymb,epsfig,float,psfrag}
\newcommand{\beq}{\begin{equation}}
\newcommand{\eequ}{\end{equation}}
\newcommand{\eeq}{\end{equation}}
\def\bea{\begin{eqnarray}}
\def\eea{\end{eqnarray}}
\def\as{\relax\ifmmode\alpha_s\else{$\alpha_s${ }}\fi}
\def \pt{\relax\ifmmode{p_t}\else{$p_t${ }}\fi}

\newcommand{\noi}{\noindent}

\def\as{{\alpha_s}}
\def\be{\begin{equation}}
\def\ee{\end{equation}}
\def\ba{\begin{eqnarray}}
\def\ea{\end{eqnarray}}

\newif\ifdtup

\def\eqal2#1{\,\vcenter{\openup1\jot
\caja   \ialign{\strut \hfil$\displaystyle{##}$&\hfil$
\displaystyle{{}##}$\hfil &$
\displaystyle{{}##}$\hfil\crcr#1\crcr}}\,}

\begin{document}

\title{
\begin{flushright}\normalsize
\vspace{0cm}
DESY 02-047 \\ 
hep-ph/0204263\\
April 2002
\vspace{1.cm}
\end{flushright}
\bf  Exclusive $B\to K^{**} \gamma$ Decays in the QCD LCSR
approach\footnote{Talk presented at the XXXVIIth Rencontres de Moriond:
Electroweak Interactions and Unified Theories, Les Arcs, France, 9-16 March 2002.}\unboldmath}

\author{A. S. Safir \\
        DESY, Deutsches Elektronen-Synchrotron, D-22603 Hamburg, Germany \\
        E-mail~: safir@mail.desy.de} 

\par \maketitle

\maketitle
\vspace{0.2truecm}

\begin{abstract}
We predict contributions of higher $K$-resonances to the radiative
rare decays $b\to s\gamma$, in the framework of the QCD sum rules on
the light cone (LCSR). 
Our calculations are restricted to the leading twist-two operators for
$K^*(892)$ and to the asymptotic wave function for the other
$K^{**}$-mesons. Using experimental data on the semileptonic $\tau\to
K^{**}\nu_{\tau}$ decays, we extract the corresponding decay constants
for vector and axial-vector $K^{**}$-mesons. We present results for
the corresponding branching ratios and  compare them with the
existing theoretical predictions.

\end{abstract}


\section{Introduction}\hspace*{\parindent}
The study of radiative decays based on the flavour-changing neutral
$b\to (s,d)+\gamma$ current transition is of crucial importance for
testing the flavour sector of the Standard Model and probing for new
physics. 
In the standard model, the short distance contribution to rare B-decays is dominated by the top quark, and long distance contributions by form factors. Precise measurements of these transition will not only provide a good estimate of the top quark mass and the CKM matrix elements $V_{td},\ V_{ts},\ V_{tb}$, but also of the hadronic properties of $B$-mesons, namely form factors which in turn would provide a good knowledge of the corresponding dynamics and more hint for the non-perturbative regime of QCD.


Experimental measurements of exclusive $B\to K^*\gamma$ branching ratios 
have been reported by the BABAR, CLEO and BELLE Collaborations, with the 
results:
\begin{eqnarray}
   10^5\,\mbox{Br}(\bar B^0\to\bar K^{*0}\gamma) 
   &=& \cases{ 4.23\pm 0.40\pm 0.22 & ~\protect\cite{BaBar}\cr
		4.55_{\,-\,0.68}^{\,+\,0.72}\pm 0.34 &
                ~\protect\cite{CLEO} \cr
		4.96\pm 0.67\pm 0.45 & ~\protect\cite{Ushiroda:2001sb}}
    \nonumber
\end{eqnarray}

and also of the inclusive rate \cite{CLEO,Ushiroda:2001sb,ALEPH}: 
\begin{eqnarray}
10^{4}Br(B \to X_{s} \gamma)=(3.22\pm 0.40)
 \nonumber
\end{eqnarray}

However, the first observation of the rare $B$-decay to the orbitally
excited strange mesons has been reported by CLEO \cite{CLEO}, and
recently by BELLE \cite{Ushiroda:2001sb}:
\begin{eqnarray}
10^{5} Br(B \to K_2^*(1430) \gamma)=\cases{
(1.66 ^{+0.59}_{-0.53} \pm 0.13)&\cite{CLEO} \cr 
(1.26 \pm 0.66\pm 0.10)&\cite{Ushiroda:2001sb}\cr}\nonumber
\end{eqnarray}
These important experimental measurements provide a crucial challenge
to the theory. Whereas many theoretical approaches have been employed
to predict the exclusive $B\to K^*(892) \gamma$ decay rate, less attention has been devoted to rare radiative $B$-decays to excited strange mesons \cite{altomari,mannel,Veseli,Faustov}. Most of these theoretical approaches rely on non-relativistic quark models\cite{altomari,mannel}, HQET \cite{Veseli} and relativistic model\cite{Faustov}. However there is a large spread between different results, due to  different treatments of the long distance effects.

In this talk we present the results of \cite{Safir}, where a systematic
analysis of the electromagnetic penguin form factor $F^{K^{**}}_1(0)$
governing the exclusive rare $B$-decays to orbitally excited
$K^{**}$-mesons was performed in the framework of the QCD sum rules
on the light cone \cite{Balitsky}.

\section{General framework}\hspace*{\parindent}
At the quark level, the rare semileptonic decay 
$b \to s  \gamma$ can be described in terms of the effective 
Hamiltonian obtained by integrating out the top quark and $W^\pm$ 
bosons: 
\begin{equation}
H_{eff} = -4 \frac{G_F}{\sqrt{2}}  V_{t s}^\ast  V_{tb}  
\sum_{i=1}^{8}  C_{i}(\mu)  O_i(\mu) \; . 
        \label{eq:he}
\end{equation}

\noi where $V_{ij}$ are the corresponding CKM matrix elements and
$G_F$ is the Fermi coupling constant. Following the notation and the convention of
ref.\cite{AlietGreub}, regarding the operator basis, one can test the model dependence of the form factors for the exclusive decay in the ratio of the exclusive-to-inclusive radiative decay branching ratio:

\begin{eqnarray}
R_{K^{**}} &\equiv& {BR(B \to K^{**}\ \gamma)\over BR(B \to X_s\
\gamma)}\simeq F^{K^{**}}_1(0)^2 \zeta (m_s, m_b,m_{K^{**}},..)\label{eqRV} 
\end{eqnarray}
\noi where $\zeta (m_s, m_b, m_{K^{**}},..)$ is a kinematic function
which can be found in \cite{Safir}. With this normalization, one
eliminates the uncertainties from the CKM matrix elements and the short distance contribution. Thus, we are left in (\ref{eqRV}) with unknown form
factors $F^{F}_1(0)$, which we will derive using QCD sum rules on the light
cone.

The starting point of our sum rule is to consider the correlation function
\begin{equation}
 i \int {dx} \ e^{i qx} \times \nonumber  <K^{**}(p,\epsilon)|T \left \{\bar{\psi}(x)
\sigma_{\mu \nu}(\sigma_{\mu \nu}\gamma_{5}) q^{\nu}  b(x) \bar{b}(0) i \gamma_{5}\psi(0)\right\}|0>\label{eq01}  
\end{equation}

\noi Hereafter we use $\psi$ as a generic notation for the field of
the light quark. The hadronic representation of (\ref{eq01}) is obtained by
inserting a complete set of states including the $B$-meson ground
state, higher resonances and the non-resonant states with $B$-meson
quantum numbers. After writing down the dispersion relation in $(p+q)^2$, 
we can separate the contribution of the $B$-meson as the pole contribution.

The possibility to calculate the correlator (\ref{eq01}) in the
region of large space-like momenta $(p+q)^2<0$ is based on the
expansion of the $T$-product of the currents in (\ref{eq01}) near the light-cone $x^2=0$ 
which is expressed through matrix elements of non-local operators,
sandwiched in between the vacuum and the meson state. These matrix
elements define the light-cone meson wave functions. We restricted our
calculations to the leading twist-2 operator for the $K^*(892)$, as in \cite{AlietBraun}, and to the asymptotic wave function for the other $K^{**}$-mesons. The latter choice is simply based on the fact that using QCD sum rules, it is impossible to get rid of the lower-lying states contributions from these higher resonances.

However, nothing is known about the corresponding $K^{**}$-decay
constants, and one has to predict them. For that, we have used recent
data \cite{PDG} on semileptonic $\tau \to (K^{**}) \nu_{\tau}$ decays \footnote{These quantities contribute to the transition rates for pseudoscalar, vector, scalar and axial vector emission.} to obtain them \cite{Safir}. We present in table \ref{tab:p1} the
corresponding decay constants. For $K^{*}_{2}(1430)$, we have constrained the  corresponding decay constant with the recent data \cite{CLEO} on $B\to K^{*}_{2}(1430)\gamma$.

Following the basic steps of the QCD sum rules on the light-cone, as
described above, and using the experimental $K^{**}$-decay constants,
we show in table \ref{tab:p1} the corresponding  form factors. 
Finally, in table \ref{tab2}  we compare our results for the ratio $R_{K^{**}}[\%]$ with
previous works \cite{altomari,mannel,Veseli,Faustov}.

\begin{table}
\addtolength{\arraycolsep}{0pt}
\renewcommand{\arraystretch}{1}
\caption[]{\it Central values of the pseudoscalar, vector, scalar and axial vector $K^{**}$-meson decay constants (in MeV) and the corresponding form factors.}\label{tab:p1}
$$
\begin{array}{|l|lllllll|}
\hline
&K^*(892) &K^*_1( 1270)  &K_{1}(1400) &K^*(1410) & K^*_{0}(1430) &K_{1}(1650) &K^*(1680)\\ 
\hline
J^P &1^-  &1^+ &1^+ &1^- & 0^+ &1^+ &1^-\\
f_{i}  &210 & 122 & 91 & 86 & 79& 86& 86 \\
F_1^{K^{**}}(0) &0.32^{+0.06}_{-0.06} &0.14^{+ 0.03}_{- 0.03} & 0.098^{+0.02}_{-0.02} &0.094^{+0.02}_{-0.02} & &0.091^{+0.02}_{-0.02} &0.091^{+0.02}_{-0.02}\\\hline
\end{array}
$$
\end{table}

\begin{table}
\begin{center}
\caption{\it Comparison of our results for the ratio  $R_{F}[\%]$ with
previous works.} 
\label{tab2}
\begin{tabular}{|c|ccccc|}
\hline   & & & $R_{F}[\%]$ & &\\
\hline
Meson      & ref.\cite{Safir}   &  ref.\cite{Veseli} &  ref.\cite{mannel} &  ref.\cite{altomari} &  ref.\cite{Faustov}   \\
\hline
$K$     & \multicolumn{4}{c|}{forb.} & \\
$K^{*}(892)$ &$10.0^{+ 4.0}_{- 4.0}$ &
$16.8\pm 6.4$ & $3.5-12.2$ & 4.5 &$15^{+ 3}_{- 3}$\\
$K^{*}(1430)$ &  \multicolumn{5}{c|}{forb.} \\
$K_{1}(1270)$ &$2.0^{+ 0.8}_{- 0.8}$ &
$4.3^{+ 1.6}_{- 1.6}$ & $4.5-10.1$ & forb./6.0 &$1.5^{+0.5}_{-0.5}$\\
$K_{1}(1400)$ & $0.9^{+ 0.4}_{- 0.4}$ &
$2.1^{+ 0.9}_{- 0.9}$ & $6.0-13.0$ &forb./6.0 &$2.6^{+0.6}_{-0.6}$\\
$K^{*}_{2}(1430)$ &$5.0^{+ 2.0}_{- 2.0}$ &
$6.2^{+ 2.9}_{- 2.9}$ & $17.3-37.1$ & 6.0&$5.7^{+ 1.2}_{- 1.2}$\\
$K^{*}(1680)$ &$0.7^{+ 0.3}_{- 0.3}$ &
$0.5^{+ 0.2}_{- 0.2}$ & $1.0-1.5$ & 0.9&\\
$K_{2}(1580)$ & &
$1.7^{+ 0.4}_{- 0.4}$ & $4.5-6.4$ & 4.4& \\
$K(1460)$    & 
\multicolumn{4}{c|}{forb.}& \\
$K^{*}(1410)$ &$0.8^{+ 0.4}_{- 0.4}$ &
$4.1^{+ 0.6}_{- 0.6}$ & $7.2-10.6$ & 7.3&\\
$K^{*}_{0}(1950)$ & 
\multicolumn{4}{c|}{forb.}& \\
$K_{1}(1650)$ &$0.8^{+ 0.3}_{- 0.3}$ & $1.7^{+ 0.6}_{- 0.6}$ & not given& not given &\\
\hline
\end{tabular}
\end{center}
\end{table}

\section{Summary}\hspace*{\parindent}
Motivated by the first observation of the radiative decay $B \to
K^{*}_{2}(1430)\gamma$, we have investigated rare radiative $B$ decays
to orbitally excited $K^{**}$-mesons. 
First, we have presented an alternative method of calculating the
transition form factors and related decays using the QCD sum rules on
the light-cone. For that, We have extracted the unknown $K^{**}$-decay
constants using the recent data \cite{PDG} on semileptonic $\tau \to K^{**} \nu_{\tau}$ decays.

For $K^{*}_{2}(1430)$, we have constrained the  corresponding decay
constant with the recent data \cite{CLEO} on $B\to
K^{*}_{2}(1430)\gamma$. We find that if $f_{K^{*}_{2}(1430)}=
(140-180) \ MeV$, a substantial fraction $(3.0-7.0)\%$ of the
inclusive $b\to s \gamma$ branching ratio goes into the
$K^{*}_{2}(1430)$ channel, in a good agreement with recent CLEO data
\cite{CLEO}. Our prediction for the $B \to K^{*}(892)\gamma$ branching
fraction yields to $(6.0-14.0)\%$, also in good agreement with the
experimental data \cite{CLEO,BaBar,Ushiroda:2001sb}.

In order to make comparison of our results with previous calculations, we have tabulated our results together with results of \cite{altomari}, \cite{mannel}, \cite{Veseli}  and \cite{Faustov}  in table \ref{tab2}.
As far as decays into higher $K$-resonances are concerned, our results
are in general in much better agreement with \cite{Veseli} than
\cite{altomari} and \cite{mannel}, apart from the
$K^{*}(1410)$-channel where the difference is more significant. Finally, it should be noticed that the theoretical uncertainties in our light-cone sum rules are the wave functions and the decay constants of the  $K^{**}$-mesons. The accuracy of our calculation can be substantially improved by taking into account the wave functions of twist-3 and twist-4 for the $B \to K^{*}(892)\gamma$ decay, and going beyond the asymptotic form for the other decay modes. To reduce the uncertainties on the $K^{**}$-mesons decay constants, one can determine them independently using QCD sum rules for the two-point correlator of the corresponding currents.

\section*{Acknowledgments}
\hspace*{\parindent} I would like to thank the Organizers of the
XXXVIIth Rencontres de Moriond for their financial support. I express
my gratitude to A. Ali for critical reading of the manuscript.


\end{document}